# Calibration of Shielded Microwave Probes Using Bulk Dielectrics

K. Lai, W. Kundhikanjana, M.A. Kelly, Z.X. Shen

Department of Applied Physics and Geballe Laboratory for Advanced Materials
Stanford University, Stanford, CA 94305

Abstract

A stripline-type near-field microwave probe is micro-fabricated for microwave impedance microscopy. Unlike the poorly shielded co-planar probe which senses the sample tens of microns away, the stripline structure removes the stray fields from the cantilever body and localizes the interaction only around the focused-ion beam deposited Pt tip. The approaching curve of an oscillating tip toward bulk dielectrics can be quantitatively simulated and fitted to the finite-element analysis result. The peak signal of the approaching curve is a measure of the sample dielectric constant and can be used to study unknown bulk materials.

Near-field microscopy investigates the interaction between a sub-wavelength probe and matter to obtain the electromagnetic information beyond the diffraction limit.[1] In order to quantitatively analyze the localized probe-sample interaction, the far-field radiation and any non-tip part of the sensor in the near field should be properly shielded. This is particularly important at low frequencies, such as in the microwave regime,[2] with free-space wavelength above millimeters, or the near-field signal inevitably contains parasitic interaction between the probe and environment. Unfortunately, due to the difficulty of fabricating micron-sized coaxial probes, tip shielding has not been extensively addressed by groups actively working on near-field microwave microscopy.[3-8] The non-local stray field contribution not only compromises the spatial resolution but also complicates the quantitative analysis of tip-sample interaction. Calibration of the microscope output using bulk samples [9-11] is therefore difficult, if not totally impossible.

In an earlier work, we demonstrated the fabrication of atomic-force-microscopy (AFM) compatible near-field probes and the operation of the microwave impedance microscope (MIM).[12,13] Unlike the conventional metal-wire probe backed by a relatively massive resonator,[3,4,7,8] the cantilever probe is scalable and significantly miniaturized for better resolution. For better sensitivity, metal traces rather than lossy doped Si lines [14,15] were patterned on nitride cantilevers. However, the co-planar structure provides insufficient shielding to the transmission lines carrying microwave signals, compromising the ability to perform localized and quantitative studies.[13] In this paper, we present the result of a stripline cantilever-based microwave probe, in which the Al traces are sandwiched between silicon nitride layers and shielded by grounded metal layers, therefore localizing the probe-sample interaction down to micron-sized volumes. Using finite-element analysis (FEA), we simulate the approaching curve toward bulk dielectric samples. After calibration with dielectric standards, the MIM can be used to determine the dielectric constant of materials including high-k insulators.

Fig. 1(a) shows the scanning electron micrograph (SEM) of the shielded cantilever probe. The layout of the handle Si chip is similar to the earlier co-planar probe detailed in Ref. [12]. As seen in Fig. 1(b), the layer structure of the cantilever, from bottom up, consists of 2μm of low pressure chemical vapor deposited (LPCVD) $Si_3N_4$, 1μm thick Al traces (two 10μm lines spaced by 60μm), 2μm plasma-enhanced chemical vapor deposited (PECVD) $Si_3N_4$, and 0.5μm Al front metal. A thin layer of Al is then blank coated onto the back of the cantilever to complete the stripline structure. The cantilever bending due to the composite materials is 10°~15° at the room temperature, acceptable for scanning purposes. Near the probe end, a 2μm PECVD $SiO_2$ pedestal is formed. A 2μm via hole in the center of this pedestal is drilled through both $SiO_2$ and upper $Si_3N_4$ layers and reaches the Al trace. Focused-ion beam (FIB) deposited Pt is then back-filled into this via hole to form the tip electrode, as shown in the schematic in Fig. 1(b) and SEM in Fig. 1(c). A tip radius less than 100nm, which sets the spatial resolution, is routinely achieved. The other Al trace, through a different via connection to the front metal, forms a second electrode surrounding the tip (not shown).[12] In this study, this guard electrode is held at ground and only reflection signals from the tip are detected.

The stripline structure ensures that the tip-sample interaction only occurs near the micron-sized Pt tip, with a dimension six orders of magnitude smaller than the wavelength at our working frequency of 1GHz. The near-field interaction is therefore quasi-static (static in space while oscillating in time) in nature and can be modeled as lumped elements.[13] We employ the 2D axisymmetric model in a commercial FEA program COMSOL 3.4 to calculate the tip-to-ground capacitance.[16] The geometry of the tip, as sketched in Fig. 1(d), is determined from the SEM picture. Fig. 1(d) also shows the mesh generated for the finite-element modeling. The program then calculates the quasi-static potential distribution and the effective tip impedance, including the reactive (mostly capacitive) and resistive components. From the simulation, accurate determination of the tip height and the diameter of its base, which merely alters the result several microns away from the sample, is not essential. The modeling also confirms that the stray coupling due to an unshielded transmission line (0.5mm × 10μm in area, 10μm away from the tip-sample contact) is in the order of 1fF ($10^{-15}$F), much larger than the pertinent tip-sample capacitance of 0.01 ~ 0.1fF. We therefore conclude that shielding of the electrode is crucial for localizing the tip-sample interaction and any quantitative studies.

The MIM system setup is shown in Fig. 2(a). The electronics is identical to the one detailed in Ref. [13], in which the reflected signal is routed to 50Ω transmission lines, suppressed by common-mode cancellation, amplified by RF amplifiers, and finally demodulated by the mixer. For dielectric samples, the phase of the mixer reference is adjusted such that the output only occurs in one channel (MIM-C). In order to remove the slow varying background introduced by temperature and other electronic drifts, an oscillating voltage is applied to the z-piezo to vibrate the probe at 400Hz and the MIM signal at the same frequency is detected by a lock-in amplifier.[17] Since this frequency is much below the cantilever's mechanical resonance (~150kHz), the entire probe moves as a rigid body and its motion well controlled. In particular, as the tip approaches the sample, the motion immediately pauses whenever hard contact occurs, as shown schematically in the inset of Fig. 2(b). From this point, continuing tip-sample approaching further deflects the cantilever and eventually stops the tip.

The MIM-C signals when a tip approaches a polished $SiO_2$ crystal (dielectric constant $\varepsilon_r = 4.5$) with constant oscillating amplitude of 100nm are shown in Fig. 2(b). The lock-in amplifier simultaneously records both the first (400Hz) and second (800Hz) harmonics of the mixer output. As described above, the tip moves in a truncated sinusoidal fashion when partially stopped by the sample. Therefore, the $1^{st}$ harmonic signal first increases for closer spacing, peaks when partially in contact, and then decreases and vanishes in the end. Assuming a total gain of 110dB, the simulated MIM-C signals, also decomposed into harmonics of the tapping frequency, are plotted in Fig. 2(b) for comparison. Almost perfect agreement between the experimental data and the modeling can be achieved for these approaching curves, with fitting parameters being the geometric parameters in Fig. 1(d) and the system gain. Indeed, such quantitative calibration of the system response is only possible for the shielded probe. For the same approaching process, the co-planar probe [12,13], with insufficient shielding, senses the presence of a sample even when it is 50μm away, as shown in Fig. 2(c). And the signal does not vanish after the tip is stuck on the surface. On the contrary, the stripline probe starts interacting with the sample only when the spacing is less than 3~4μm, comparable to the size of the Pt tip. The shielded probe is thus much superior for microwave imaging applications.

The approaching curve, especially the peak $1^{st}$ harmonic MIM-C signal bears important dielectric information about the sample. For the co-planar probes with a strong parasitic effect from the cantilever body, only relative information within the sample can be obtained. For the shielded stripline probes, the demodulated lock-in signal is now an absolute measurement of the tip-sample interaction. Due to the compliance of the cantilever, tip damage is not significant, as confirmed by only 10% increase of the signal after many approaching experiments. It is also verified that the orthogonal MIM-R output stays at zero for all lossless specimens. As a result, one can perform quantitative study on bulk dielectric materials using calibrated system response from standard samples. In Fig. 3 we show the results (tapping amplitude 70nm) on several bulk dielectrics – thick film thermal $SiO_2$ ($\varepsilon_r = 3.9$), polished $SiO_2$ ($\varepsilon_r = 4.5$), $CaF_2$ ($\varepsilon_r = 6.8$), and sapphire ($\varepsilon_r \sim 10$) crystals.[18] The measurement error bar represents the systematic uncertainty due to non-linearity of the z-piezo and the moderate topography of the polished sample surface. The solid line in Fig. 3 corresponds to the expected peak MIM-C signal modeled by FEA, assuming 110dB system gain and 70nm tapping amplitude. Excellent agreement between the modeling and experimental data is again obtained for these insulators. Using this fitting curve, $\varepsilon_r$ of other dielectrics such as organic polymers, high-k dielectrics, can be measured. We include another data point on a piece of (110) $TiO_2$ crystal in Fig. 3. The low frequency dielectric constant of $TiO_2$ crystals is usually around 100 and differs along different crystal axis.[19] Due to the monopole geometry of the probing fields, the effective dielectric constant, measured to be $\varepsilon_r \sim 140\pm20$, is an average response sensed by the tip.

In summary, we have fabricated shielded cantilever-based probes for microwave impedance microscopy. The shielding of the metal lines removes the stray signal from the cantilever body and enables quantitative analysis of the tip-sample interaction. Using standard lock-in techniques, the approaching curve of the MIM-C channel output is accurately described by the lumped-element FEA simulation. The system response can be calibrated using bulk dielectric samples and $\varepsilon_r$ of unknown materials measured. We expect to further implement the tapping operation, compatible with conventional AFM tapping mode, for quantitative near-field microwave imaging.

The research is funded by the seed grant in Center of Probing the Nanoscale (CPN), Stanford University, with partial support from a gift grant of Agilent Technologies, Inc. and DOE contract DE-FG03-01ER45929-A001. CPN is an NSF NSEC, NSF Grant No. PHY-0425897. The cantilevers were fabricated in Stanford Nanofabrication Facility (SNF) by A.M. Fitzgerald and B. Chui in A.M. Fitzgerald & Associates, LLC, San Carlos, CA.

**Figure Caption:**

Figure 1: (Color online) (a) SEM picture of the shielded cantilever probe. (b) Schematics of the layer structure of the cantilever body (top) and the tip region (bottom), as indicated by the dashed lines in (a). (c) Zoom-in SEM image near the tip area, showing the front Al layer, the $SiO_2$ pedestal, and the FIB Pt tip. A second Al electrode surrounding the tip is also seen in the picture. (d) Left – Dimensions of the Pt tip used throughout this paper for FEA simulation. Right – Mesh near the tip apex generated by the COMSOL program.

Figure 2: (Color online) (a) Schematics of the MIM setup. The shielded probe vibrates at low frequencies due to an AC excitation on the z-piezo. The reflected signal is suppressed by the common-mode cancellation through a directional coupler (D), amplified by RF amplifiers (A), and demodulated by an IQ-mixer (M). The modulated signal is detected by a lock-in amplifier. (b) Simulated (solid lines) and measured (empty squares and circles) MIM-C signal as a function of the average tip-sample spacing. Both the 1$^{st}$ and 2$^{nd}$ harmonics of the output are plotted. The inset shows the actual tip motion, which is truncated when hard contact occurs. (c) Comparison of MIM-C signals between the co-planar and stripline probes as a function of average tip-sample spacing. Both curves are scaled by the maximum signal for clarity.

Figure 3: (Color online) Peak MIM-C signal in tip-sample approaching as a function of the relative dielectric constant. Data from several insulators (symbols), as well as the FEA modeling result (solid line), are plotted. The data point at the upper-right is taken on a piece of $TiO_2$ crystal and its effective dielectric constant estimated to be 140±20 from the measurement.

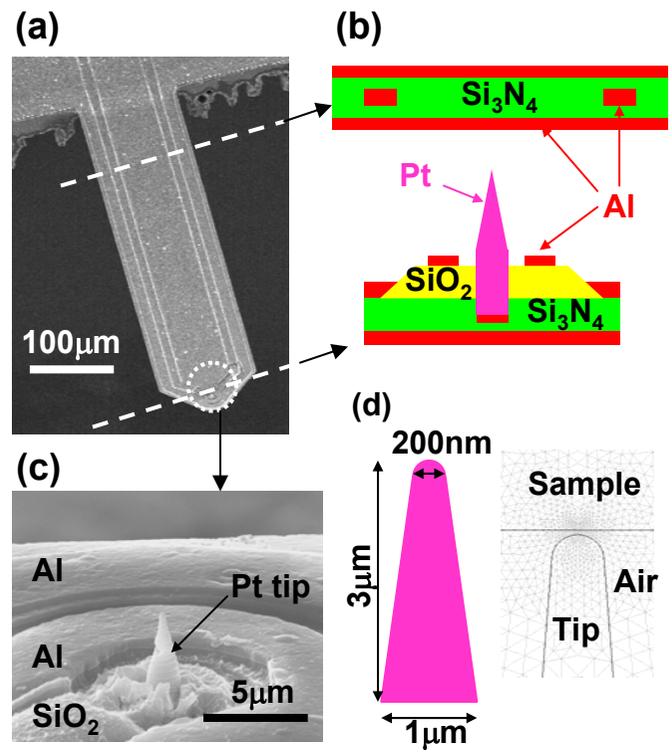

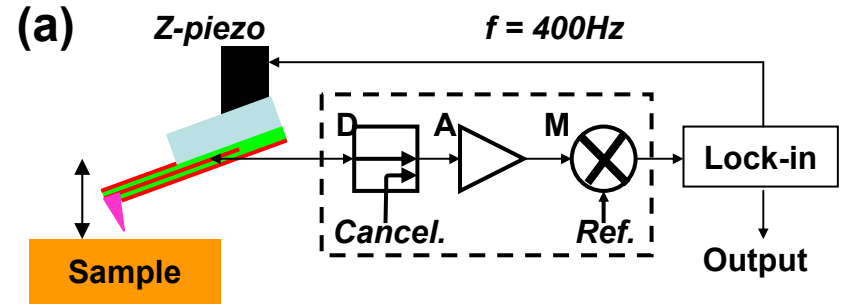
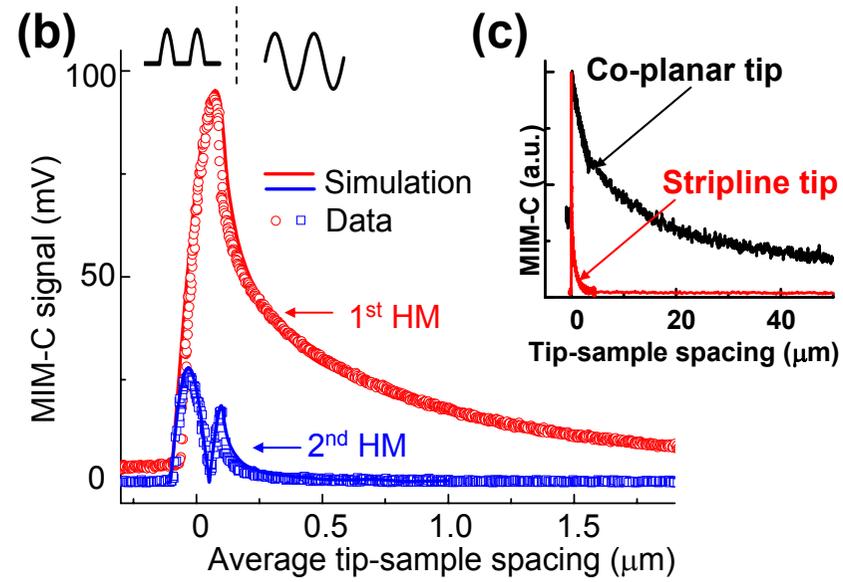

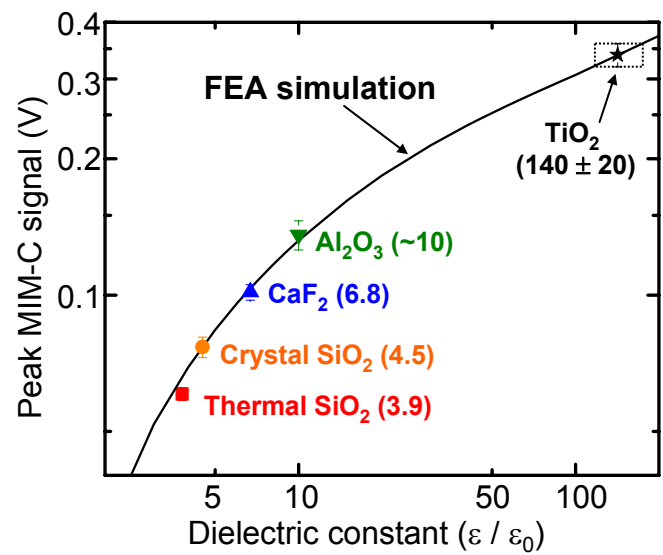